# Advances in quantum optomechanical control based on cavity cold-atom systems*

LI Baijun[1], LU Tianxiang[2], WANG Jie[3], JING Hui[3, 4]

1. Research Center for Quantum Physics, Huzhou University, Huzhou 313000, China
2. College of Physics and Electronic Information, Gannan Normal University, Ganzhou 341000, China
3. College of Science, National University of Defense Technology, Changsha 410073, China
4. Synergetic Innovation Center for Quantum Effects and Applications, Key Laboratory of Low-Dimensional Quantum Structures and Quantum Control of Ministry of Education, Department of Physics, Hunan Normal University, Changsha 410081, China

**Abstract** Cavity optomechanics represents a frontier research direction in quantum science and technology, centered on the control and exploration of interactions between electromagnetic fields and mechanical motion. Rooted in quantum mechanics, this field realizes efficient coupling and precise manipulation of optical fields and mechanical vibrations through the structural design and precise regulation of optical microcavities and mechanical oscillators, thereby revealing and harnessing novel quantum phenomena. In recent years, cavity optomechanics has evolved into an interdisciplinary frontier integrating quantum optics, condensed matter physics, and quantum precision measurement, exhibiting profound application in both fundamental physics research and quantum information science. With advancements in technologies such as nanomanufacturing and laser cooling, optomechanical interactions have been successfully demonstrated in various experimental systems. Among these platforms, cavity cold-atom systems stand out as one of the ideal platforms for implementing quantum optomechanics. Featuring exceptional environmental isolation, long quantum coherence times, and strong light-matter interaction, these systems provide a crucial testbed for exploring strong and even ultrastrong optomechanical coupling effects as well as rich nonlinear quantum phenomena. This paper reviews the recent progress in optomechanical control and manipulation based on cavity cold-atom systems. We first outline the fundamental principles of standard cavity optomechanical systems. Then,

* The paper is an English translated version of the original Chinese paper published in Acta Physica Sinica. Please cite the paper as: **LI Baijun[1], LU Tianxiang[2], WANG Jie[3], JING Hui[3, 4] , Advances in quantum optomechanical control based on cavity cold-atom systems .** *Acta Phys. Sin.*, 2026, 75(11): 110301. DOI: 10.7498/aps.75.20260289

we describe the experimental realization of linear and nonlinear optomechanical couplings in cold-atom systems. After that, we focus on the representative applications of this platform in high-precision quantum sensing, quantum memory, and the preparation of macroscopic nonclassical states. Finally, we give an outlook and challenges in this field.

# 1 Introduction

In the 17th century, Kepler's observations of comet trajectories provided evidence for the momentum characteristics of photons. However, due to the weak intensity of radiation pressure, significant physical effects induced by light pressure were not effectively observed or investigated in the early stages. The emergence of laser technology in the 1960s[1], coupled with continuous advancements in micro- and nanofabrication techniques[2], promoted the development of high-quality optical microcavities. These cavities confine photons within microscopic spaces for cyclic reflection, significantly enhancing radiation pressure effects. This advancement accelerated the rapid development of cavity optomechanics, establishing it as a highly promising research direction in quantum optics and precision measurement. Cavity optomechanics has demonstrated important applications in two core research areas: 1) achieving efficient coupling and control between optical and mechanical degrees of freedom via radiation pressure, which holds promise for controlling mechanical motion at the quantum scale and preparing non-classical states; and 2) enabling ultra-high sensitivity detection of physical quantities such as weak forces, displacement, mass, and acceleration, thereby providing an excellent platform for quantum precision measurement. Over the past two decades, cavity optomechanics has achieved significant progress. The realization of ground-state cooling of mechanical oscillators[3-6] marked the entry of cavity optomechanics into the quantum regime. This achievement serves not only as a major milestone in fundamental research of quantum mechanics but also lays the foundation for the implementation of various quantum technologies. The observation of optomechanically induced transparency[7,8] has provided crucial support for the development of fast and slow light control and optical storage technologies. Furthermore, the theoretical prediction and experimental verification of non-classical states, including single-photon states[9], squeezed states[10-12], entangled states[13-16], and superposition states[17,18], have further strengthened the application basis of cavity optomechanics in quantum information technology. In the field of ultra-high sensitivity sensing, the gravitational wave detection experiments honored by the 2017 Nobel Prize in Physics relied partially on cavity optomechanics principles. Additionally, the system has made important progress in weak force sensing[19], displacement sensing[20], mass sensing[21], acceleration sensing[22], and magnetometry[23], demonstrating broad application prospects. Benefiting from the efficient coupling of mechanical oscillators with various physical systems such as microwaves and atoms, cavity optomechanical systems can serve as quantum information interfaces, enabling quantum state conversion between optical and microwave bands[24]. Given its immense application potential in fundamental physics, precision metrology, and quantum information, cavity optomechanics has been listed by *Nature* as one of the 23 milestones in the history of optics. To date, the core interactions of cavity optomechanics have been verified in experimental systems using two different approaches. The first is the "top-down" approach, which includes Fabry-Pérot (FP) cavity systems[25], whispering gallery mode microdisk cavity systems[26] and microsphere cavity systems[27], photonic crystal systems[28-30], levitated systems[5,6,31-33], and optical waveguide systems[34]. The second is the "bottom-up" approach, such as cavity cold atom systems[35-40]. Different types of optomechanical systems are suitable for different application scenarios, and their system parameter ranges also vary. Among these, optomechanical systems based on cavity cold atoms possess unique advantages. Their optomechanical coupling strength can be flexibly tuned, allowing them to approach or even reach the strong coupling regime. Moreover, the mechanical oscillators are immune to thermal noise, providing an ideal platform for in-depth exploration of quantum optomechanics and the development of novel quantum devices.

The field of cavity cold atoms is one of the frontier research directions at the intersection of quantum optics, cavity quantum electrodynamics, and atomic and molecular physics. Its core lies in the coupling between cold atom ensembles and optical microcavities, a research framework that has greatly expanded understanding of the nature of light-matter interactions. By trapping atom ensembles within high-quality factor optical microcavities and combining the strong confinement characteristics of microcavities for light fields with the technical advantages of laser cooling and atomic manipulation, researchers have achieved efficient coupling between atom ensembles and cavity fields, as well as precise control of atomic gases. Leveraging the unique physical properties of cavity cold atom systems, they have become ideal experimental platforms for studying quantum coherence, quantum simulation, and quantum many-body physics. The physical basis of cavity-atom systems originates from cavity quantum electrodynamics. Early cavity quantum electrodynamics focused on studying atomic radiation characteristics in the presence of boundaries[41], whereas cavity cold atoms began with the exploration of cavity-enhanced interactions. Experimental findings revealed that trapping cold atoms in high-quality optical microcavities significantly enhances light-atom coupling strength[42]. By adjusting parameters such as optical cavity properties, atomic density, and laser field configurations, one can flexibly achieve weak, strong, or even ultra-strong coupling between atoms and light fields. In most experimental scenarios, the effective coupling strength between the atom ensemble and the cavity field is proportional to the square root of the number of atoms[43-46]. Using such systems, cavity-enhanced atom-photon entanglement has been realized[47]. For laser-driven cavity atom ensembles, the local cavity field intensity experienced by a single atom depends on the spatial distribution of other atoms[48], which can generate effective long-range or global interatomic interactions[49]. Recently, long-range exchange interactions mediated by optical cavities were observed in strontium atom ensembles within high-finesse optical cavities, leading to one-axis twisting dynamics[50]. In transversely driven cavity-atom ensemble systems, such long-range interactions can induce self-organization phase transitions, causing atoms to arrange in a checkerboard pattern that maximizes the coupling of scattered light to the cavity mode[51,52]. Breakthroughs in laser cooling and atom trapping technologies[53-55] have provided key support for the rapid development of the cavity cold atom field. These technologies enable the preparation of ultracold atom ensembles and even Bose-Einstein condensates (BEC)[56-58] in highly controlled states. The combination of BEC with optical microcavities has further enriched the physical connotations and research dimensions of cavity cold atom systems. Ultracold atoms possess characteristics such as long coherence times, high atomic density, and strong interactions with cavities, providing favorable conditions for exploring new quantum states such as quantum phase transitions. For example, coupling a laser-driven BEC with the cavity vacuum field can realize quantum phase transitions between superfluid and self-organized phases[59,60]. This system can also be used to simulate quantum many-body models with tunable long-range interactions, such as the Dicke model and its quantum phase transitions[61-64]. Notably, in the laser-driven cavity BEC model, strong coupling between the cavity field and the collective motion modes of the BEC can be achieved, building an important bridge for extended research in cavity optomechanics. In cavity BEC systems, the collective motion modes of atoms can be equivalent to mechanical oscillators. Their coupling with the cavity field provides a new perspective for exploring the essence of optomechanical interactions and novel control methods. Currently, cavity optomechanics based on cavity BEC systems has been verified in multiple experimental platforms[35-40]. Compared with traditional solid-state optomechanical systems, cavity optomechanical systems based on ultracold atoms exhibit irreplaceable advantages due to the unique quantum propertie. First, the BEC collective density excitations or the collective center-of-mass motion of ultracold atoms, serving as mechanical modes, exist in the quantum ground state or at extremely low temperatures. This effectively suppresses thermal noise, laying the foundation for quantum optomechanical effects. Second, the coupling strength between photons and mechanical oscillators can be tuned over a wide range by flexibly adjusting system parameters, reaching the strong or even ultra-strong coupling regime. This allows mechanical displacements induced by single photons to be distinguished from quantum zero-point fluctuations, making optomechanical control at the single-photon level possible. Due to the advantages, numerous important effects have been theoretically predicted or experimentally realize, such as the preparation of macroscopic or microscopic non-classical states[65-67], ground-state cooling of mechanical modes[38], high-sensitivity rotation sensing[68], efficient quantum state transfer[69], quantum memory[70], and multistable oscillations[71]. This system provides a solid foundation for the cross-integration of cavity optomechanics and cold atom physics, as well as for conducting various frontier application research. This article reviews the latest progress in quantum optomechanical control research based on cavity cold atom systems.

# 2 Basic Principles of Cavity Optomechanical Systems and Experimental Realization Based on Cavity Ultracold Atom Systems

## 2.1 Basic Principles of Cavity Optomechanical Systems

The core physical essence of cavity optomechanical systems is the use of radiation pressure to achieve coupling and control between optical and mechanical degrees of freedom. The most typical system is the FP cavity optomechanical system, whose typical structure is shown in Figure 1. This system mainly consists of a standard FP optical microcavity formed by a pair of parallel high-reflectivity mirrors, where the distance between the two mirror surfaces is the cavity length. In this system, the left mirror is fixed to ensure the stability of the cavity structure. The right mirror is movable and can undergo reciprocating vibration along the cavity axis. It also possesses high reflectivity, enabling near-total reflection of the intracavity light field, minimizing light field energy loss, and enhancing the interaction between light and the mechanical oscillator. When a laser beam is input into the system, it enters the optical cavity through the fixed mirror and undergoes multiple cyclic reflections between the two parallel mirrors, forming a stable standing wave. When photons inside the cavity strike the movable mirror, they exchange momentum with the mirror surface. Driven by radiation pressure, the right mirror undergoes simple harmonic vibration along the cavity axis. The movement of the movable mirror causes a change in the cavity length, which leads to a shift in the resonant frequency of the cavity mode. Therefore, under the action of radiation pressure, a dependency relationship forms between the optical mode frequency and the position of the mechanical oscillator, ultimately achieving effective coupling between the optical field degree of freedom and the mechanical degree of freedom.

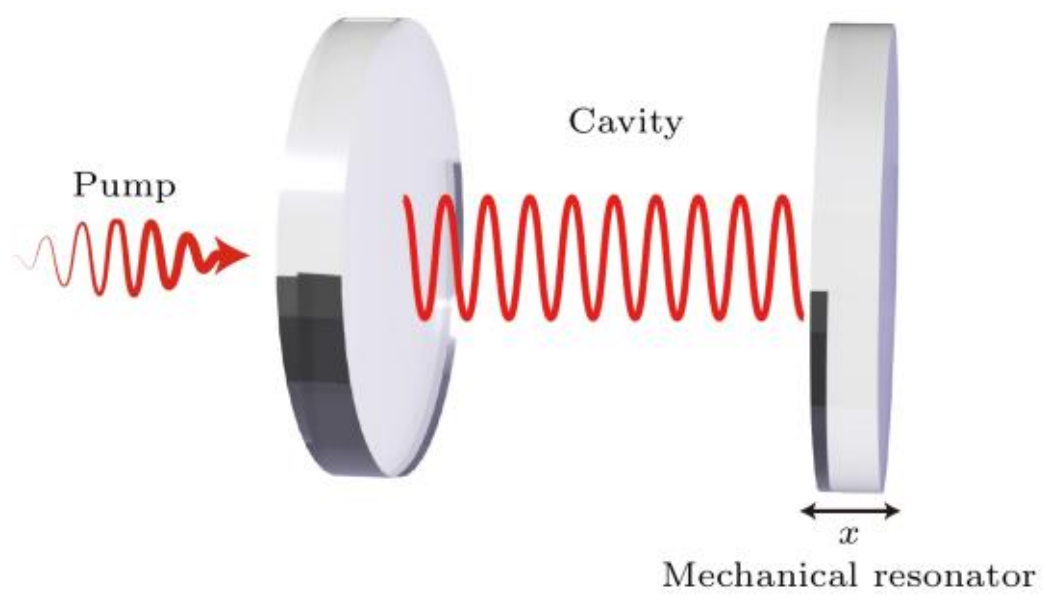

**Fig. 1 A typical schematic diagram of cavity optomechanical system (Fabry-Pérot (FP) optomechanical system).**

When the right movable cavity mirror is stationary, the effective length of the cavity is $L$. To form a stable standing wave for the incident laser within the cavity, the standing wave resonance condition must be satisfied. Specifically, the effective cavity length must equal an integer multiple of the half-wavelength of the laser. Based on the relationship, the resonance frequency of the cavity mode is given by $\omega_{\mathrm{c}} = n\pi c/L$, where $n$ is an integer and $c$ is the speed of light in a vacuum. Under the influence of optical radiation pressure, the right movable mirror displaces, thereby altering the effective cavity length. The cavity length changes from the initial $L$ to $L + x$, where $x$ represents the displacement of the movable mirror along the cavity axis. This slight change in cavity length causes a shift in the resonant frequency of the cavity mode. By substituting the changed cavity length into the aforementioned frequency formula, the resonant frequency of the cavity is obtained as

$$\omega_{\mathrm{c}}(x) = \frac{n\pi c}{L+x} = \omega_{\mathrm{c}}\frac{1}{1+x/L}. \qquad (1)$$

Due to the weak radiation pressure, the displacement $x$ is much smaller than the initial cavity length $L$ and can be treated as a small quantity relative to $L$, i.e., $x \ll L$. We expand the oscillator displacement in a Taylor series, neglect higher-order terms, and retain only the first-order term. The cavity frequency is thus further simplified as[72]

$$\omega_{\mathrm{c}}(x) \approx \omega_{\mathrm{c}}(1 - x/L). \qquad (2)$$

It is evident that the cavity mode resonance frequency depends on the displacement of the movable cavity mirror. The mechanical displacement modulates the cavity field resonance frequency, thereby demonstrating the control of optical degrees of freedom by mechanical degrees of freedom. Consequently, the Hamiltonian of the entire system can be expressed as

$$\begin{aligned} H = &\quad \hbar\omega_{\mathrm{c}}(x)a^{\dagger}a + \hbar\omega_{\mathrm{m}}b^{\dagger}b \\ = &\quad \hbar\omega_{\mathrm{c}}a^{\dagger}a + \hbar\omega_{\mathrm{m}}b^{\dagger}b - \hbar\frac{\omega_{\mathrm{c}}}{L}a^{\dagger}ax. \end{aligned} \tag{3}$$

Here, $a$ $(a^{\dagger})$ and $b$ $(b^{\dagger})$ denote the creation (annihilation) operators for the optical microcavity and the mechanical oscillator, respectively. The displacement of the mechanical oscillator is expressed as $x = x_{\mathrm{zpf}}(b^{\dagger} + b)$, where $x_{\mathrm{zpf}}$ represents the zero-point fluctuation of the mechanical oscillator. Based on these definitions, the typical form of optomechanical interaction can be written as $\hbar g a^{\dagger}a(b^{\dagger} + b)$, where $g$ is the single-photon coupling strength. Its typical range is $10^{-3}$—$10^{6}\mathrm{Hz}$[72]. This type of optomechanical interaction, characterized by the dependence of the cavity mode frequency on oscillator displacement, is commonly referred to as dispersive optomechanical interaction. In cavity optomechanical systems, dissipative optomechanical coupling constitutes another important form of nonlinear coupling, in addition to the dispersive type. Its core principle involves influencing the loss of optical modes through the displacement of the mechanical oscillator, thereby achieving coupling between the optical field and mechanical degrees of freedom[73-75]. In traditional cavity optomechanical systems, dissipative optomechanical interaction has been applied in studies such as cooling of mechanical oscillator and quantum state preparation and control[76,77]. Typically, the strength of dissipative optomechanical coupling is much smaller than that of dispersive optomechanical coupling.

It is important to note that achieving strong single-photon coupling remains a significant challenge for traditional optomechanical systems. In contrast, cavity cold-atom optomechanical systems provide an ideal platform for realizing strong single-photon coupling. Furthermore, optomechanical interaction is inherently a typical nonlinear interaction, which complicates the analysis of system dynamics. Linearization approximation is currently a common approach for addressing such problems. This method decomposes the relevant operators of the optical field and mechanical oscillator into the sum of a steady-state average value and a small fluctuation term with a zero mean. By neglecting higher-order cross terms of the fluctuations, the process of solving the system dynamics equations is significantly simplified. This facilitates an effective analysis of the fundamental physical properties of the system[7].

## 2.2 Experimental Implementation of Optomechanical Models in Cavity

## Ultracold Atom Systems

This section focuses on experimental schemes and progress in realizing optomechanical interactions $\hbar g a^{\dagger} a(b^{\dagger}+b)$ between the collective motion of quantum gases and single-mode optical microcavities. In this context, the cavity field primarily senses a single collective motion of the atomic ensemble, and this collective motion acts as a mechanical oscillator. To date, optomechanical nonlinearity has been realized in cavity ultracold atom systems using two distinct experimental schemes. First, the collective center-of-mass motion of the ultracold atomic ensemble serves as the mechanical oscillator to establish strong coupling with the cavity field. Second, collective density excitations in a Bose-Einstein condensate (BEC) act as the mechanical oscillator to achieve strong coupling with the cavity field.

The center-of-mass motion of an ultracold atomic ensemble serves as the mechanical oscillator. In 2007, Gupta et al.[78] utilized a Fabry-Pérot (FP) cavity to couple with an ultracold rubidium ($^{87}$Rb) atomic gas, achieving linear optomechanical coupling. They successfully observed refractive nonlinear effects and bistability in the system. In this experiment, the drive applied by the intracavity optical field displaced the trapped atoms. This modulation adjusted the position-dependent coupling strength between the atomic ensemble and the cavity mode. In 2008, Murch et al.[36] prepared an ultracold atomic gas consisting of $10^5$ $^{87}$Rb atoms in a high-precision FP cavity (Figure 2(a)). They verified that the collective motion of a macroscopic ultracold atomic ensemble can serve as a mechanical resonator in quantum optomechanics. The collective displacement sensed by the cavity is equivalent to the center-of-mass motion of the atomic ensemble, confirming the existence of linear optomechanical coupling. In this system, a single collective motion degree of freedom of the atomic ensemble couples only to a single mode of the cavity field. The coordinate and momentum operators for the collective motion of the atomic ensemble were defined. Benefiting from the advantage of cooling ultracold atoms to the quantum ground state, this mechanical oscillator effectively overcame thermal noise interference. This enabled the observation of quantum measurement backaction effects. In the experiments, the atoms were not trapped in a single potential well. Instead, they were confined in a one-dimensional optical lattice composed of multiple potential wells, occupying approximately 300 adjacent lattice sites. In 2011, Schleier-Smith et al.[38] investigated sideband cooling for a single collective motion mode of an atomic ensemble in an FP cavity-coupled atomic ensemble model. The minimum thermal conduction and the observed cooling rate agreed well with the optomechanical model. Beyond single-oscillator cavity optomechanical systems, cavity ultracold atom systems can be extended to multi-mechanical-oscillator cavity optomechanical systems. In 2015, Spethmann et al.[39] trapped two $^{87}$Rb cold atomic ensembles within an FP cavity (Figure 2(b)), successfully constructing a dual-mechanical-oscillator cavity optomechanical system. In this experiment, they confined approximately 900 atoms in two adjacent superlattice potential wells. Deep superlattice barriers effectively suppressed atom exchange between the two atomic ensembles. The center-of-mass motions of the two atomic clouds served as independent mechanical oscillators. These formed linear couplings with the optical field, constituting a cavity optomechanical system composed of two mechanical oscillators. Based on this system, researchers achieved coherent interactions between two addressable, spatially separated, near-ground-state mechanical oscillators within the driven optical cavity. This provided an important experimental foundation for the control of multi-oscillator quantum optomechanics.

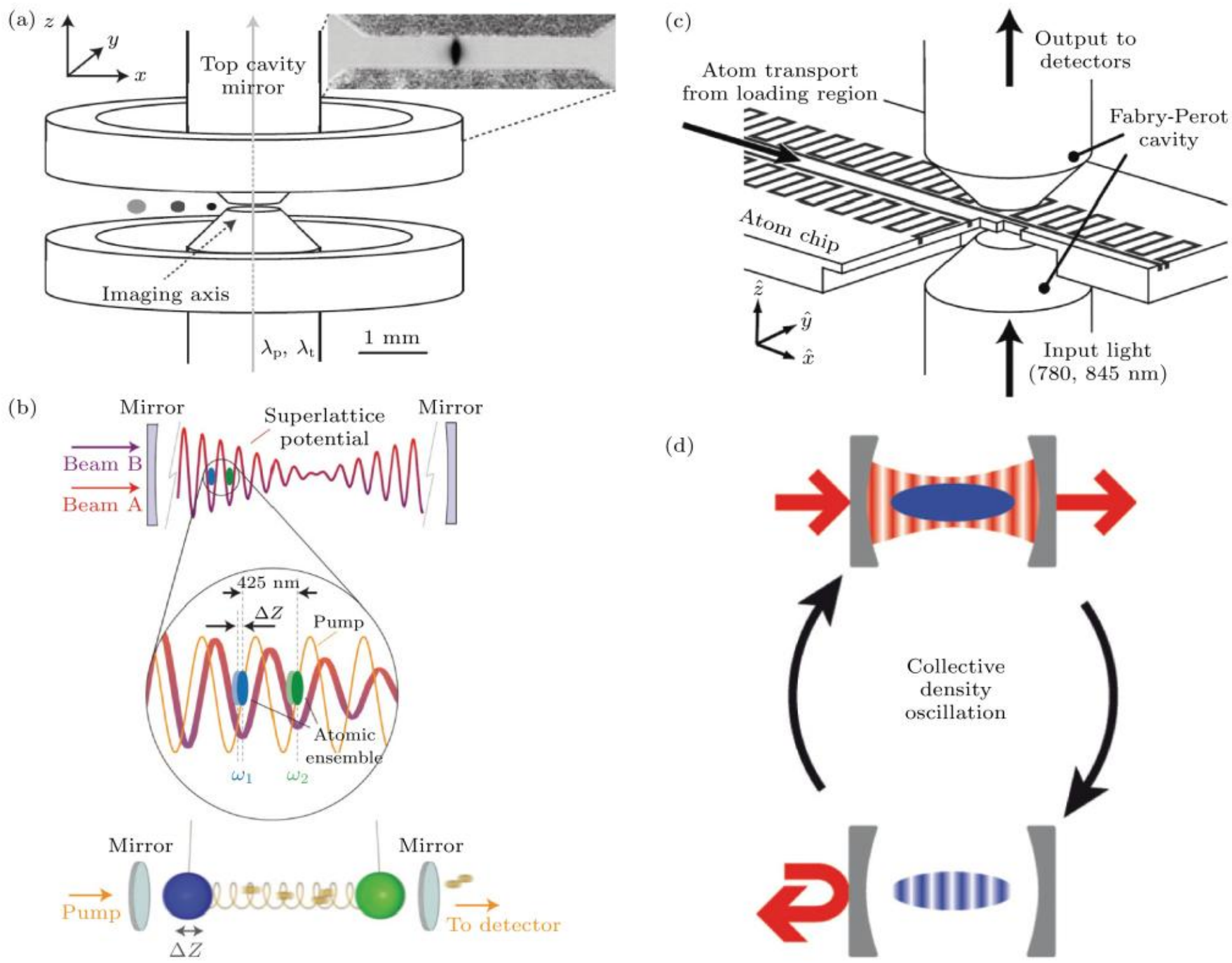

**Fig. 2 Key experiments for realizing cavity optomechanical system (COMs) using cavity ultracold atom. (a) $^{87}Rb$ ultracold atoms ensemble is confined in a FP cavity. The center-of-mass motion of the atoms ensemble serves as a mechanical oscillator. Reproduced with permission from Ref.[36]. (b) Two $^{87}Rb$ atoms ensembles are confined in a FP cavity. The centers of mass of each ensemble form a mechanical oscillator, constituting a dual-mechanical-oscillator COMs. Reproduced with permission from Ref.[39]. (c) $^{87}Rb$ ultracold atoms ensemble is localized in a FP cavity, while an atomic chip is employed for atomic transport and positioning. This system enables realization of linear and quadratic COMs. Reproduced with permission from Ref.[40]. (d) A BEC formed from $^{87}Rb$ atoms is coupled to a FP cavity, where the collective density oscillations of the BEC act as a mechanical oscillator to realize a linear COMs. Reproduced with permission from Ref.[35].**

In addition to the linear optomechanical coupling, ultracold atomic systems can also realize quadratic optomechanical coupling[79-81]. In 2010, Purdy et al.[40] implemented a tunable cavity optomechanical system (Fig. 2(c)), which was constructed by integrating a microfabricated atom chip with a Fabry-Pérot (FP) cavity. In the experiment, nanometer-precision positioning of cold atoms was achieved by combining the atom chip and optical potential trap. The atomic ensemble was not confined to a single lattice site; however, most atoms occupied no more than two adjacent lattice positions. By positioning the atomic gas and precisely controlling its center-of-mass position along this axis, the system simultaneously realized both linear and quadratic optomechanical coupling. Researchers systematically investigated the influence of tunable coupling on optomechanical bistability and frequency shifts. They also completed the quantitative characterization of optomechanical effects under quadratic coupling, providing significant experimental support for the study of nonlinear quantum optomechanics.

Collective density excitations in a Bose-Einstein condensate (BEC) serve as mechanical oscillators. For cavity-BEC systems, experimental schemes for realizing linear optomechanical forces have been explored. In 2008, Brennecke et al.[35] first realized a linear cavity optomechanical system within a cavity-BEC setup (Fig. 2(d)). This scheme utilized the collective density excitations of the BEC as mechanical oscillators, effectively suppressing thermal excitations of the oscillator by leveraging the quantum properties of the BEC. In the experiment, the cavity field coupled to a collective density excitation mode of the BEC. This density excitation mode matched the spatial distribution of the cavity mode, thereby achieving strong optomechanical coupling. The experiment employed a BEC composed of typically $1.2 \times 10^5$ $^{87}$Rb atoms coupled to an ultra-high-finesse optical FP cavity. A weak driving field was incident along the cavity axis to drive the coupled dynamics between the cavity field and the BEC. Under large detuning conditions, spontaneous emission from the atoms was effectively suppressed. In momentum space, the BEC initially macroscopically occupied the zero-momentum state. Through the absorption and stimulated emission of cavity photons, atoms were scattered into momentum states $\pm 2\hbar k$ distributed symmetrically along the cavity axis. Matter-wave interference occurred between the macroscopically occupied zero-momentum component and the recoil momentum components. This led to periodic spatial modulation of the condensate density, with a modulation frequency equal to four times the recoil frequency[82]. When scattering effects from higher-order momentum modes are neglected, the condensate field operator can be approximated as a linear combination of the zero-momentum mode and its first two side modes. In this regime, the system dynamics are described by a linear optomechanical Hamiltonian. Here, the phonon mode corresponds to the collective excitation of the recoil momentum modes, with an oscillation frequency equal to four times the recoil frequency.

# 3 Optomechanical Control Based on Cavity Cold Atom Systems

In conventional "top-down" optomechanical systems, optomechanical control has demonstrated significant advantages across multiple research fields. In 2012, the Dong et al. [83] realized a mechanical dark mode using optomechanical control, as shown in Fig. 3(a). The dark mode fraction (the ratio of particles in the dark mode to the total particles in both bright and dark modes) reached 99%. They also achieved optical field transfer between two optical modes using this dark mode. In the field of nonreciprocal transmission, the Dong et al. realized a non-magnetic optical isolator in 2016 through optomechanical control[84] (Fig. 3(b)). They subsequently expanded this work to include optomechanical circulators, directional amplifiers, and nonreciprocal frequency conversion[85,86]. In 2021, they achieved cooling and amplification effects of mechanical oscillations in cascaded optomechanical systems[87]. Furthermore, researchers confirmed the existence of synthetic gauge fields in a single optomechanical cavity[88]. In laser physics research, Grudinin et al.[89] predicted the phonon laser. Jing et al.[90], Zhang et al.[91], and the Wu et al. [92] combined optomechanical control with exceptional points to achieve theoretical exploration and experimental verification of ultra-low-threshold phonon laser (Fig. 3(c)). Jiang et al.[93] and Li et al.[94] further extended this research to nonreciprocal phonon lasers. In nonlinear dynamics research, the Zhang et al. observed chaos-induced stochastic resonance in optomechanical systems[95] and subsequently observed mechanical solitons[96]. The Lü et al. predicted chaos with an ultra-low driving threshold[97] and further investigated squeezing-enhanced optomechanics[98], as well as mechanical squeezing and phonon blockade effects via optomechanical interaction[12,99]. Recently, Sang et al.[100] achieved spatial optical nonreciprocity in the telecommunication band using metamaterials and optomechanical control. Li et al.[23,101] realized high-precision magnetic field sensing. Tang et al.[102] achieved synchronous frequency locking of two kHz mechanical modes within an optomechanical crystal cavity.

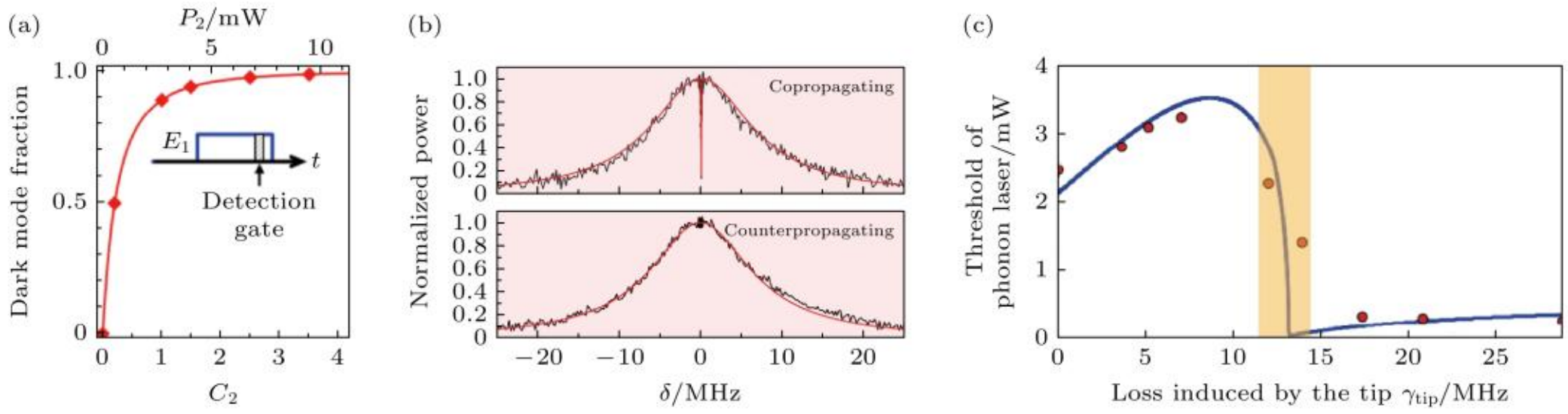

**Fig. 3 Top-down optomechanical systems: dark modes, nonreciprocity, and ultralow-**

**threshold phonon lasers. (a) Dark mode fraction versus optomechanical cooperativity. Reproduced with permission from Ref.[83]. (b) Emission power spectra of clockwise and counterclockwise signals versus dutuning. Reproduced with permission from Ref.[84]. (c) Threshold power of phonon laser versus dissipation. Reproduced with permission from Ref.[91].**

Due to the long lifetime of Bose-Einstein condensate (BEC) condensates and the advantages of cavity-based BEC optomechanical systems, such as strong optomechanical nonlinearity, low thermal noise, and multi-parameter tunability, numerous novel classical and quantum effects have been theoretically explored in these systems. The following section reviews a series of effects generated by optomechanical control in recent cavity cold atom systems.

## 3.1 Rotational Quantum Sensing and Quantum Storage in Cavity Ring BEC Optomechanical Systems

Persistent currents in ring-trapped superfluids[103] serve as an important platform for studying quantum circulation phenomena, including gyroscopes[104], cosmology simulations[105], and phase slips[106]. Therefore, measuring the rotational states of ring BECs is significant and has substantial implications for multiple fields of physics. Typically, the angular momentum information of a rotating BEC eigenstate is encoded in its phase, manifested as the winding number of topological vortices. However, existing methods for detecting the BEC winding number require observing atoms within the ring via absorption imaging, which destroys the condensed state of the BEC. In 2021, Kumar et al.[68] applied the principles of cavity optomechanics to provide a method for sensing the rotation of ring BECs in situ and in real time with minimal destructiveness. This scheme avoids fully destructive measurements. Furthermore, they demonstrated that light can actively manipulate rotating matter waves through optomechanical entanglement with persistent currents. The model they considered is shown in Figure 4(a), where an atomic BEC is confined in a ring trap located at the center of an optical cavity[107]. The BEC can be rotated using various methods, such as optical stirring[103], radio-frequency fields[108], and quenching techniques[109], thereby endowing the BEC with a winding number $L_{\mathrm{p}}$. In this context, the winding number $L_{\mathrm{p}}$ corresponds to its macroscopic orbital angular momentum quantum number. These two quantities are positively proportional and share the same physical meaning; thus, the corresponding orbital angular momentum is $\hbar L_{\mathrm{p}}$[68]. Two frequency-degenerate driving beams, carrying superimposed orbital angular momenta of $\pm l\hbar$, are injected into the system to probe the BEC information. This coherent superposition has been proven to form an angular lattice around the cavity axis[110]. Due to the presence of the optical lattice, some atoms in the condensate undergo coherent Bragg scattering[111], scattering from the rotational state with winding number $L_{\mathrm{p}}$ to states $L_{\mathrm{p}} \pm 2nl$. When the dipole potential is weak, only first-order diffraction needs to be considered, resulting in states $L_{\mathrm{p}} \pm 2l$. Analogous to assumptions in BEC optomechanical simulations, the atomic field operator can be approximated as the sum of the original mode ($L_{\mathrm{p}}$) and two side modes ($L_{\mathrm{p}} \pm 2l$). Due to the weak dipole potential, the number of atoms in the side modes is small, so the mode with winding number $L_{\mathrm{p}}$ remains macroscopically occupied. Under this assumption, the system can be transformed into an optomechanical model, with the two side modes acting as mechanical oscillators at different frequencies. Simultaneously, the BEC winding number is encoded in the frequencies of these mechanical oscillators, and the intracavity optical field is modulated by these two mechanical oscillation frequencies. Therefore, atomic information can be inferred from the optical field information. Homodyne measurement of the cavity output field, combined with noise spectrum analysis, clearly reveals the expected peaks of the mechanical oscillator frequencies, from which the BEC winding number $L_{\mathrm{p}}$ can be extracted. Furthermore, the system sensitivity was calculated using the error propagation formula. As shown in Figure 4(b), the researchers demonstrated the function of orbital angular momentum number $l$ versus sensitivity. They found that sensitivity increases with the value of $l$ as more optical lattice points interact with the BEC. Under practical parameter conditions, the maximum sensitivity of the cavity optomechanical control method for detecting BEC rotation relative to a stationary laboratory is three orders of magnitude higher than current experimental levels[112] and is comparable to theoretical predictions based on destructive measurements[104]. Finally, the researchers also demonstrated that light can entangle the two rotating matter wave side modes.

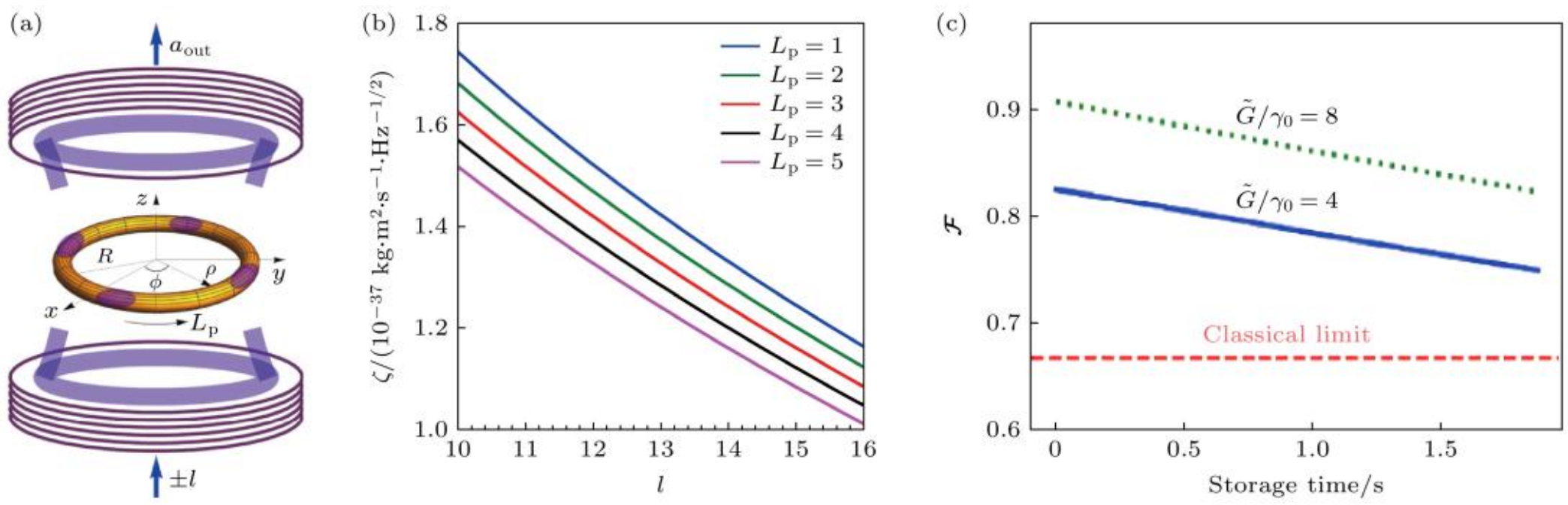

**Fig. 4 Rotation quantum sensing and quantum memory based on cavity ring BEC optomechanical system. (a) An atomic BEC confined in a toroidal trap located at the center of an optical cavity. Reproduced with permission from Ref.[107]. (b) Rotation sensitivity as a function of orbital angular momentum number (OAM) *l*. Reproduced with permission from Ref.[107]. (c) Retrieval fidelity as a function of storage time for the storage and retrieval of OAM superposition states. Reproduced with permission from Ref.[70].**

In addition to quantum sensing, the optomechanical control based on cavity ring-shaped Bose-Einstein condensates (BECs) can realize photonic orbital angular momentum quantum memories with long storage times. Optical quantum storage devices[113,114] can store and retrieve optical quantum states as required, constituting an essential component of quantum information science. Leveraging the long lifetime advantage of ring-shaped BECs[115,116], optical quantum states can be mapped onto the persistent current modes of rotating BECs. This approach provides an effective pathway for realizing long-lived orbital angular momentum quantum memories. In 2026, Daloi et al.[70] proposed a cavity ring-shaped BEC quantum storage platform based on cavity optomechanical control for storing and reading photonic states encoded with orbital angular momentum. The model considered by the researchers is similar to that in Figure 4(a), where a BEC trapped in a ring-shaped potential well is placed within a Fabry-Pérot (FP) cavity. A signal beam and a control beam are input into the cavity to drive the system. Both fields are prepared as coherent superpositions of Laguerre-Gaussian modes carrying orbital angular momentum $\pm l\hbar$, forming a ring-shaped optical lattice around the cavity axis. Similar to the rotational sensing analysis by Kumar et al.[68], when considering only first-order Bragg scattering, the Hamiltonian of this system can be expressed as a two-mechanical-oscillator optomechanical model, where the mechanical modes correspond to the two side modes of the BEC. Following standard optomechanical memory protocols[117], the researchers achieved the storage and retrieval of single-photon Fock states, coherent superpositions and entangled two-photon states with orbital angular momentum. Furthermore, the system demonstrated high-fidelity retrieval. For instance, the fidelity for coherent superpositions of orbital angular momentum states reached 90% (Figure 4(c)), while the fidelity for entangled two-photon states reached 80%.

### 3.2 Preparation of Nonclassical States in Cavity One-Dimensional Cigar-Shaped BEC Optomechanical Systems

Nonclassical sources are vital components of quantum technologies[118-120]. Schrödinger cat states and single-photon sources (states) are typical representatives of nonclassical sources.

Schrödinger cat states refer to coherent superpositions of two or more macroscopically distinguishable states, representing a unique feature of quantum systems. These quantum states hold significant importance for quantum technologies. They allow for conceptual exploration of the boundary between classical and quantum regimes[121] and provide key fault-tolerant quantum resources for applications such as quantum information processing and quantum sensing[122]. Due to their significance, Schrödinger cat states at the mesoscopic scale have been extensively explored in systems such as photons[123-125], ions[126], and atoms[127]. Notably, researchers prepared Schrödinger cat states in a 16 μg mechanical resonator[18]. Optomechanical systems[17,128,129] also demonstrate superior capabilities for realizing both microscopic and macroscopic Schrödinger cat states. However, achieving ultrastrong single-photon coupling remains a significant challenge in current cavity optomechanical systems, thereby hindering the generation of distinguishable Schrödinger cat states[130]. Cavity BEC optomechanical systems, benefiting from large optomechanical coupling and low thermal phonon noise, represent ideal platforms for realizing distinguishable Schrödinger cat states. In 2012, Zhang et al.[65] placed a three-dimensionally trapped atomic BEC inside a single-mode unidirectional ring high-$Q$ optical cavity, realizing inverted cavity optomechanics. In this model, the roles of the optical field and the matter-wave field are reversed: the matter-wave field acts as the optical field, while the optical field acts as the mechanical oscillator. Since the mechanical oscillator is simulated by the optical field, the system is unaffected by thermal noise. The researchers found that this system could efficiently prepare nonclassical states of the BEC and allow for their nondestructive characterization and Wigner function reconstruction. In 2023, Li et al.[66] investigated large-scale mechanical superposition states and tunable optical multi-component cat states in a standard cavity BEC optomechanical model, leveraging its advantages. As shown in Figure 5(a), the researchers considered a cavity BEC system where a cigar-shaped atomic BEC is trapped in an FP cavity. The collective density excitations of the BEC are equivalent to a mechanical oscillator, forming a cavity optomechanical system with the optical field. Based on this system, the researchers explored macroscopic mechanical superposition states and microscopic optical field Schrödinger cat states. Leveraging strong optomechanical nonlinearity, the maximum amplitude of coherent states in Schrödinger cat states generated in cavity BEC optomechanical systems is approximately three orders of magnitude higher than that obtained in solid-state systems, reaching a distinguishable level. Figure 5(b) displays the Wigner function of the mechanical mode at a fixed time. Two main peaks in phase space are clearly visible, representing the vacuum state and the coherent state, respectively. Distinct interference fringes appear between the two peaks. This quantum interference provides clear evidence of macroscopically distinct superposition components. Furthermore, the researchers demonstrated the impact of tunability in the BEC on optical cat states, showing that adjusting the s-wave scattering length of the system can generate optical cat states with different components.

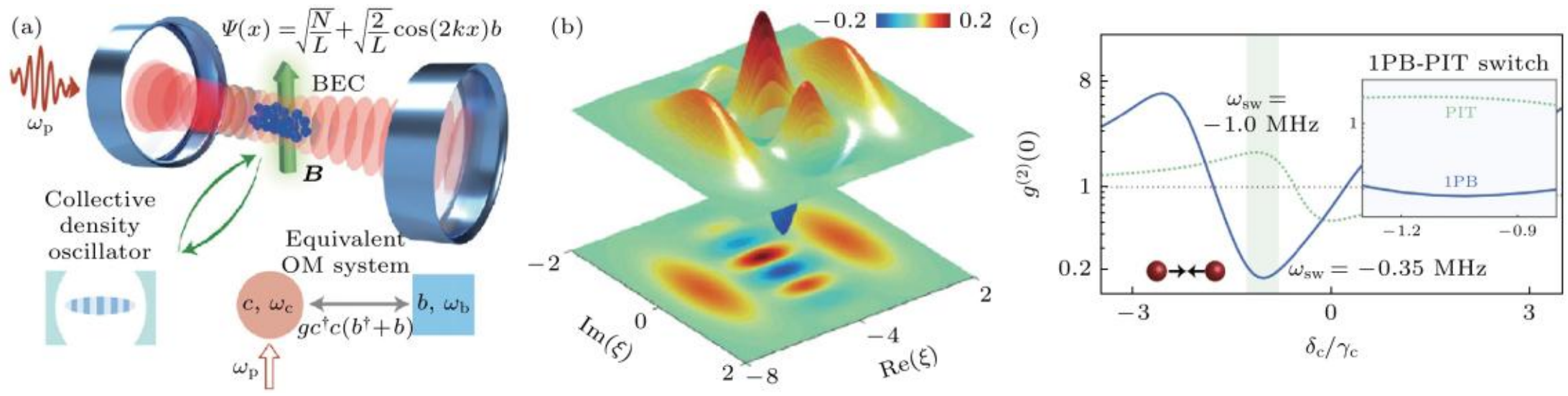

**Fig. 5 Schrödinger cat states and photon blockade in a cavity optomechanical system based on cavity cigar-shaped BEC. (a) The cigar-shaped atomic BEC is confined in an optical cavity. The collective density excitation of the BEC can be modeled as a mechanical oscillator, thereby equivalent to an optomechanical system. Reproduced with permission from Ref.[66]. (b) Wigner function of the mechanical mode in the cavity BEC system. Reproduced with permission from Ref.[66]. (c) Second-order correlation function versus detuning for different *s* scattering lengths. Reproduced with permission from Ref.[67].**

Photon blockade, a phenomenon in which the probability of generating a second photon in a nonlinear cavity is nearly zero after the generation of the first, represents a crucial method for producing single-photon sources. To date, photon blockade has been experimentally realized in cavity or microwave quantum electrodynamics (QED) systems[131-134]. Theoretically, numerous schemes involving nonlinear cavities have been proposed to achieve photon blockade[135-140]. This phenomenon has also been extensively studied in cavity optomechanical systems[9,141-143]. However, conventional photon blockade typically requires strong optomechanical nonlinearity, which must exceed cavity dissipation. The significant optomechanical nonlinearity inherent in cavity Bose-Einstein condensate (BEC) optomechanical systems provides favorable conditions for generating photon blockade. In 2024, Zhai et al.[67] investigated photon blockade in a cavity BEC system and demonstrated its advantages. The system considered is similar to that shown in Figure 5(a), where a BEC is trapped within an optical microcavity and driven by a weak field. The study reveals that photon blockade can occur in this system under strong optomechanical coupling. Furthermore, under specific parameters, the cavity BEC system can simultaneously exhibit single-photon and two-photon blockade effect. Moreover, this system offers substantial tunability: by adjusting the s-wave scattering length between BEC atoms, one can achieve transitions between photon-induced tunneling and blockade. Figure 5(c) illustrates the second-order correlation function as a function of detuning under different scattering parameters, demonstrating the feasibility of switching between photon blockade and photon-induced tunneling. Additionally, the study demonstrated the enhancing effect of scattering parameters on photon blockade.

Beyond the aforementioned significant effects, optomechanical control in cavity cold atom systems plays a vital role in various other aspects. In 2011, Dong et al.[71] demonstrated a system composed of a two-component BEC dispersively interacting with a FP optical cavity, exhibiting both few-photon-level optical multistability and matter-wave multistability involving macroscopic atomic numbers. Also in 2011, Schleier-Smith et al.[38] investigated cavity sideband cooling of individual collective motional modes of an atomic ensemble in an FP cavity atomic ensemble system. In 2012, Singh et al.[69] demonstrated quantum state transfer between a BEC and an optomechanical oscillato. Theoretical verification has also been achieved for simulating cavity optomechanical systems using spin ensembles. In 2010, Brahms et al.[144] theoretically constructed a cavity optomechanical system consisting of $N$ hydrogen-like atoms and a high-finesse optical microcavity. They predicted a series of physical phenomena, including cavity-spin bistability, spin optomechanical squeezing, and spin coherent amplification and damping. In 2011, Jing et al.[145] systematically explored the cavity optomechanical properties of antiferromagnetic BECs. By investigating how to describe the system using a single-rotor model in a spin-1 BEC and a high-$Q$ factor optical cavity, they revealed an analog relationship between the torsional oscillator and the single-cavity mode in the form of quadratic optomechanical coupling.

# 4 Conclusion

Ultracold atoms or BECs offer advantages such as long coherence times, low temperatures, and multi-parameter tunability. Cavity ultracold atom optomechanical systems, constructed by coupling these atoms with high-quality optical cavities, exhibit unique superiority: 1) The optomechanical nonlinearity is broadly tunable from weak to strong regimes; 2) The system naturally operates within the low-noise quantum cavity optomechanical regime, with extremely low thermal noise levels; 3) The coupling forms between ultracold atoms and optical cavities provide additional degrees of freedom for optomechanical control. These advantages lay a solid foundation for realizing a series of important quantum optomechanical effects, particularly in physical scenarios that explicitly require strong nonlinearity. Meanwhile, the system's natural robustness against thermal noise provides an ideal platform for realizing noise-resilient quantum effects. Based on this, this paper systematically reviews recent research progress on important physical effects related to optomechanical interaction in cavity ultracold atom systems, including photon blockade, preparation and control of macroscopic non-classical state superpositions, quantum sensing, and quantum storage. The field of cavity ultracold atom optomechanics still holds vast space for development and exploration potential. For instance, the strong optomechanical nonlinearity of the system can be utilized to break existing limits of optomechanical entanglement[146] and to realize thermal-noise-robust quantum squeezing effects and phonon laser effects[89,94]. Related research outcomes can provide important theoretical and experimental support for the development of quantum technology.

Although significant progress has been made in quantum optomechanics based on cavity cold atom systems, this field still faces certain opportunities and challenges. First, while single-mode cavity cold atom optomechanical systems have been extensively explored, there remain many gaps in multimode quantum optomechanics and its interdisciplinary research with quantum many-body physics. Second, integrating cavity cold atom optomechanical systems with quantum information applications, such as achieving high-fidelity quantum state transfer, quantum memory, and quantum network interfaces, represents an important direction for future development. Although ultracold atom systems offer the advantage of low thermal noise, technical noise and atom loss persist in experiments, imposing higher requirements on high-precision quantum control. With the continuous development of theoretical schemes and experimental techniques, cavity cold atom optomechanical systems are expected to play an increasingly important role in fundamental physics research and quantum science and technology.